\documentclass[12pt,a4paper]{article}
\usepackage[utf8]{inputenc}
\usepackage{a4wide}
\usepackage{amsmath}
\usepackage{cite}
\usepackage{xcolor}
\usepackage{amssymb}
\usepackage{bm}
\usepackage{amsfonts}
 \usepackage{adjustbox}
\usepackage{booktabs}
 \usepackage{makecell}
 \usepackage{multirow}
\usepackage{pdflscape}
\usepackage{textcomp}
\usepackage{gensymb}
\usepackage{graphicx}
\graphicspath{{figures/}}
\usepackage{diagbox}
\usepackage{pifont}
\usepackage{subcaption}
\usepackage{array}
\usepackage{enumerate}
\usepackage{caption}
\usepackage{tabulary}
\usepackage{booktabs}
\usepackage{stackengine}
\usepackage{mathtools}
\usepackage{enumerate}
\usepackage[unicode]{hyperref}
\usepackage{graphicx, color}
\usepackage{hyperref}
\usepackage{mathbbol}
\usepackage{amssymb}
\usepackage{appendix}
\usepackage{verbatim}
\usepackage{geometry}
\usepackage{makecell}
\usepackage{listings}
\usepackage{mathtools}
\usepackage{dsfont}
\usepackage{accents}
\usepackage{longtable}
\usepackage{colortbl}
\newcommand*{\belowrulesepcolor}[1]{%
  \noalign{%
    \kern-\belowrulesep
    \begingroup
      \color{#1}%
      \hrule height\belowrulesep
    \endgroup
  }%
}
\newcommand*{\aboverulesepcolor}[1]{%
  \noalign{%
    \begingroup
      \color{#1}%
      \hrule height\aboverulesep
    \endgroup
    \kern-\aboverulesep
  }%
}
\definecolor{light-gray}{gray}{0.9}
\definecolor{lighter-gray}{gray}{0.95}
\allowdisplaybreaks
\numberwithin{equation}{section}
\newcommand{\be}{\begin{equation}}
\newcommand{\ee}{\end{equation}}
\def\1{\mathbf{1}}

\def\2{\mathbf{2}}
\def\3{\mathbf{3}}

\def\k{\kappa}

\DeclareMathOperator{\im}{Im}

\makeatletter
\g@addto@macro\bfseries{\boldmath}
\makeatother
\begin{document}
\title{
\begin{flushright}
% \ \\*[-80pt] 
\begin{minipage}{0.3\linewidth}
\normalsize
SISSA 14/2023/FISI\\*[50pt]
% IPMU19-xxxx\\*[50pt]
\end{minipage}
\end{flushright}
{\Large \bf 
On the Normalisation of the Modular Forms in Modular Invariant 
Theories of Flavour
\\*[20pt]
}}

\author{ 
\centerline{
S. T. Petcov $^{a,b}\footnote{Also at:
Institute of Nuclear Research and Nuclear Energy,
Bulgarian Academy of Sciences, 1784 Sofia, Bulgaria.}$
} 
\\*[5pt]
\centerline{
\begin{minipage}{\linewidth}
\begin{center}
$^a${\it \normalsize
INFN/SISSA, Via Bonomea 265, 34136 Trieste, Italy} \\*[5pt]
$^b${\it \normalsize
Kavli IPMU(WPI), UTIAS, The University of Tokyo, Kashiwa, 
Chiba 277-8583, Japan} 
\\*[5pt]
\end{center}
\end{minipage}}
\\*[50pt]}

\date{
\centerline{\small 
\bf Abstract}
\begin{minipage}{0.9\linewidth}
\medskip 
\medskip 
\small 
The problem of normalisation of the modular forms in 
modular invariant lepton and quark flavour models is discussed. 
Modular invariant normalisations of the modular forms are proposed. 
\end{minipage}
}
\begin{titlepage}
\maketitle
\thispagestyle{empty}
\end{titlepage}

%%%%%%%%%%%%%%%%%%%%%%%%%%%%%%%%%%%%%%%%%%%%%%%%%%%%%%%%%%%%%%%%%
%%%%%%%%%%%%%%%%%%%%%%  Introduction   %%%%%%%%%%%%%%%%%%%%%%%%%%
%%%%%%%%%%%%%%%%%%%%%%%%%%%%%%%%%%%%%%%%%%%%%%%%%%%%%%%%%%%%%%%%%

\section{Introduction}

 The main new feature of the modular invariance approach 
to the fundamental flavour problem in particle physics
proposed in \cite{Feruglio:2017spp}  
\footnote{For a review of the flavour problem see, e.g.,
Ref. \cite{Feruglio:2015jfa}. A brief discussion of the earlier 
proposed ``solutions'' of the flavour problem  and their drawbacks  
is given, e.g., in Ref.~\cite{Novichkov:2021evw}.
}, 
is that the elements of the Yukawa coupling 
and fermion mass matrices in the Lagrangian of the theory are
modular forms of a certain level \(N\), 
which are functions of a single complex scalar field
-- the modulus  \(\tau\) -- and, as like the fermion (matter) fields,  
have specific transformation properties 
under the action of the inhomogeneous (homogeneous) 
modular group $\Gamma \equiv PSL(2,\mathbb{Z})$
($\Gamma^\prime \equiv SL(2,\mathbb{Z})$).  
The Yukawa couplings 
and the fermion mass matrices depend also on limited number of 
constant parameters (see further).
The theory is assumed to be invariant under the whole modular group.
In addition, both the  the fermion (matter) fields
and the modular forms, present in Yukawa couplings 
and the fermion mass matrices,  
are assumed to transform in representations of an 
inhomogeneous (homogeneous) finite modular group 
of level $N$, \(\Gamma^{(\prime)}_N\), $N=1,2,3,...$, which 
plays the role of a flavour symmetry.
For \(N\leq 5\), the finite modular groups \(\Gamma_N\) 
are isomorphic to the permutation groups 
\( S_3\), \( A_4\), \( S_4\) and \( A_5\)
 (see, e.g., \cite{deAdelhartToorop:2011re}), 
while the groups \(\Gamma^\prime_N\) are isomorphic to the double
covers of the indicated permutation groups,
\(S^\prime_3 \equiv S_3\), \(A^\prime_4 \equiv T^\prime\), 
\(S^\prime_4\) and \(A^\prime_5\).

In the modular flavour models, the VEV of the modulus \(\tau\), $\tau_v$, 
can be the only source of flavour symmetry breaking, 
such that no flavons are needed. 
The VEV of \(\tau\) can also be the only source of breaking of the 
generalised CP (gCP) symmetry, which can be consistently combined with 
the modular symmetry \cite{Novichkov:2019sqv}        
(see also~\cite{Baur:2019kwi}). 
When the  flavour symmetry is broken, the modular forms 
(which are holomorphic functions of $\tau_v$),  
and thus the elements of the Yukawa coupling  and 
fermion mass matrices get fixed, and a certain flavour 
structure arises. As a consequence of the modular symmetry 
in the lepton sector,  for example, 
the charged-lepton masses, the two neutrino mass squared differences,
the three neutrino mixing angles 
(in the reference 3-neutrino mixing scheme, see, e.g., 
\cite{ParticleDataGroup:2018ovx}) and the not yet 
% measured 
known absolute neutrino mass scale, 
neutrino mass ordering and the leptonic Dirac and Majorana 
CP-violation phases, are simultaneously determined  in terms of 
a limited number of coupling constant parameters
\footnote{ 
This together with the fact that they are, via the 
corresponding modular forms, also functions of 
a single complex VEV -- that of the modulus \(\tau\) --
leads to experimentally testable 
correlations between, e.g., the neutrino mass and 
mixing observables (see, e.g., \cite{Novichkov:2018ovf}).
Models of flavour based on modular invariance 
have then an increased predictive power.
}.
A  unique characteristic of the modular framework is the fact that 
fermion mass hierarchies may follow from the properties of the modular forms, 
without fine-tuning~\cite{Novichkov:2021evw}.

 The modular symmetry approach to the flavour problem 
has been widely implemented so far primarily 
in theories with global (rigid) supersymmetry (SUSY).  
Within this SUSY framework, modular invariance is 
assumed to be a feature of the K\"ahler potential 
and the superpotential of the theory.

   Bottom-up modular invariance approach to the lepton 
flavour problem has been exploited first using the groups
\(\Gamma_3 \simeq A_4\)~\cite{Feruglio:2017spp},
\(\Gamma_2 \simeq S_3\)~\cite{Kobayashi:2018vbk}, 
\(\Gamma_4 \simeq S_4\)~\cite{Penedo:2018nmg} 
and \(\Gamma_3 \simeq A_4\) again \cite{Criado:2018thu} 
\footnote{In the ``bottom-up'' constructions,  
typically the minimal form of the   K\"ahler potential is being used. 
Possible non-minimal additions to the K\"ahler potential, 
compatible with the modular symmetry, may 
affect the predictive power of the approach~\cite{Chen:2019ewa}. 
The indicated problem is the subject of ongoing research.
 }.
After these first studies, the interest in the approach grew significantly 
and various aspects of this approach were and continue
to be extensively studied
\footnote{A rather extensive list of publications on the modular invariance 
approach to the flavour problem can be found, e.g., in 
\cite{deMedeirosVarzielas:2023crv}.
}. 
 Recently, for example, a phenomenologically viable 
anomaly-free modular flavour model, which features 
also an ``axion-less'' solution of the strong CP problem, 
% that does not require the introduction of the axion field 
was constructed  in Ref. \cite{Feruglio:2023uof}. 
%
% The “minimal” phenomenologically viable modular flavour models 
% with gCP symmetry constructed so far  
% i) of the lepton sector with 
% massive Majorana neutrinos (12 observables) contain 
% 4 real constants \cite{Ding:2022nzn} or 5 real constants 
% \cite{Novichkov:2018ovf,Novichkov:2019sqv} and the complex VEV of $\tau$;
% ii) of the quark sector (10 observables) 
% contain 7 real constants and the complex VEV of $\tau$);
% while the models of lepton and quark flavours (22 observables) include  
% 13 real constants and  the complex  VEV of $\tau$
% (see, e.g., \cite{Qu:2021jdy}).

 It should be clear from the preceding discussion that the modular 
forms play a crucial role in the modular invariance approach to 
the flavour puzzle. As we have already mentioned, 
the elements of the fermion mass matrices, on which 
the successful description of the lepton and quark 
flavours depend, are modular forms of some level $N$ 
furnishing irreducible representations of a 
finite modular group \(\Gamma^{(\prime)}_N\).
It is a well known fact that the modular forms 
furnishing irreducible representations of the 
finite modular groups are determined up to a constant.
Usually these normalisation constants 
are absorbed in the constant parameters 
which multiply each modular form 
present in the fermion mass matrices. 
Since the normalisation of the modular forms 
is arbitrary, this makes the specific values of the 
constant parameters, obtained in a given model 
by statistical analysis of the
description of the relevant experimental data 
by the model, of not much physical meaning. 
Moreover, it also makes the comparison of 
models which use different normalisations  
of the modular forms ambiguous.
The problem of normalisation of the 
modular forms can be particularly acute in 
modular flavour models in which the 
charged lepton and quark mass hierarchies 
are obtained without fine-tuning  of 
the constant parameters present in the 
respective fermion mass matrices, 
where these constant parameters 
have to be of the same order in magnitude.

  In the present article we discuss 
the problem of normalisation of modular forms 
in the modular invariant theories of flavour 
and consider two possible modular-invariant 
solutions to this problem.

%%%%%%%%%%%%%%%%%%%%%%%%%%
%
\section{The Problem}
%
%%%%%%%%%%%%%%%%%%%%%%%%%%

Consider a modular-invariant  bilinear
%%%%%%%%%%%%%%%%%%%%%%%%%%%%%%%%%%%%%%%5%
\begin{equation}
  \label{eq:bilinear}
\tilde{W} = v_h \psi^c \, M(\tau)\, \psi \,,
\end{equation}
%%%%%%%%%%%%%%%%%%%%%%%%%%%%%%%%
%
where $v_h$
is a constant and 
the matter superfields \(\psi\) and \(\psi^c\) transform under 
the action of the inhomogeneous $PSL(2,\mathbb{Z}) \equiv \Gamma$
(homogeneous $SL(2,\mathbb{Z})\equiv \Gamma^\prime$) modular group as
%%%%%%%%%%%%%%%%%%%%%%%%%%%%%%%%%%%
\begin{equation}
\label{eq:psi_mod_trans}
\begin{split}
\psi\,&\xrightarrow{\gamma}\,(c \tau + d)^{-k_\psi} \rho(\tilde{\gamma})\,\psi\,, \\
\psi^c\,&\xrightarrow{\gamma}\,
(c \tau + d)^{-k^c_\psi}\rho^c(\tilde{\gamma})\,\psi^c\,.
\end{split}
\end{equation}
%%%%%%%%%%%%%%%%%%%%%%%%%%%%%%%%%%%
%
Here $\gamma$ is an element of  $\Gamma^{(\prime)}$, 
$(-\,k^{(c)})$ is the weight of the field $\psi^{(c)}$ and 
$\rho^{(c)}$ is an irreducible representation of a 
finite inhomogeneous (homogeneous) modular group of level $N$,  
$\Gamma^{(\prime)}_{\rm N}$.
We recall that for $N \leq 5$, the groups $\Gamma_{2,3,4,5}$
are isomorphic to the non-Abelian discrete symmetry groups 
$S_3$, $A_4$, $S_4$ and $A_5$, while the groups 
 $\Gamma^\prime_{2,3,4,5}$ are isomorphic to their respective 
double covers.

In flavour models based on modular invariance, 
terms of the type  in Eq. (\ref{eq:bilinear}), 
each multiplied by a Higgs doublet or a flavon fields, 
are present in the superpotential of the 
theory (see, e.g., 
\cite{Feruglio:2017spp,Penedo:2018nmg,Kobayashi:2018vbk,Criado:2018thu}). 
These terms lead to quark and charged lepton Yukawa couplings,
and, if neutrino masses are generated by 
the type I seesaw mechanism, they lead also to the 
neutrino Yukawa coupling and the Majorana mass term 
for the right-handed (RH) neutrino fields.
Alternatively, they can give also the 
Weinberg dimension five operator.  

 We will consider the simple case when  
the bilinear in Eq. (\ref{eq:bilinear}) 
is generated by couplings to Higgs fields,  
which are singlets under the action of the finite modular 
group. Although we discuss further the case of 
only the two Higgs fields present in the 
SUSY extension of the Standard Model, 
the discussion can be easily applied to the 
case of models with flavons.

The term of interest in Eq. (\ref{eq:bilinear})
appears when the corresponding Higgs  
field obtains a non-zero vacuum 
expectation value (VEV), $v_h$.
Then the Yukawa couplings give rise 
to  the quark, charged lepton and 
neutrino Dirac mass terms, 
or to the neutrino Majorana mass 
term associated with the Weinberg operator 
(in the latter case $v^2_h$ multiplies the mass matrix).
The quantity $M$ in Eq. (\ref{eq:bilinear}) 
is a generic notation for the mass matrices 
in these mass terms as well as for the mass matrix 
in the RH neutrino Majorana mass term, which can originate 
from a term in the super potential than includes neither 
flavon nor Higgs fields 
(with $v_h$ replaced by a generic constant).

The modular invariance of the term 
in Eq. (\ref{eq:bilinear}) implies  
that $M$, and thus the fermion mass matrices in the 
modular flavour models,  is a modular form 
$Y^{K}_{\rho_Y}(\tau)$ of level \(N\) and weight 
\(K \) furnishing a representation $\rho_{Y}$ of 
$\Gamma^{(\prime)}_{\rm N}$, such that,  if 
the Higgs field leading to the bilinear (\ref{eq:bilinear}) 
is a $\Gamma^{(\prime)}_{\rm N}$ singlet with modular weigh  \((-\,k_H)\), 
we have \(K \equiv k_\psi + k^c_\psi +k_H\) and 
$\rho_{Y} \otimes \rho^{c} \otimes \rho \supset \mathbf{1}$,
$\mathbf{1}$ being the singlet representation
of $\Gamma^{(\prime)}_{\rm N}$. Thus, 
$M = v_h\,\alpha\, Y^{K}_{\rho_Y}(\tau)$, 
where $\alpha$ is a constant. 

  In realistic modular flavour models 
$M$ is a matrix involving more than one modular 
forms and thus more than one constant.
If, for example, $\psi$ furnishes a $\mathbf{3}$
representation of $S^\prime_4$ and 
$\psi^{c}$ represent three singlet representations of 
$S^\prime_4$ having different weights, 
$M$ is a $3\times 3$ matrix involving at least three different 
modular forms, each transforming  as triplet representation of 
$S^\prime_4$, and thus involves at least three different constant 
parameters. 

To be more specific, consider the case of 
$\psi$ being the three quark doublet superfields, 
$\psi = Q = (Q_1, Q_2, Q_3)$, and 
$\psi^c$ being the three singlet up-type or 
down-type quark superfields, $\psi^c = q^c = (q^c_1, q^c_2, q^c_3)$. 
Let us assume further that 
$Q$ furnishes the triplet representation of 
$S^\prime_4$, ${\bf 3}$, and carry weight $k_Q$, 
$q^c_{1,2,3}$ transform as the 
``hatted'' singlet ${\bf \hat{1}}$ of $S^\prime_4$ 
\footnote{See Ref. \cite{Novichkov:2020eep} 
for the technical details related to this discussion.
}
but have different weights $k_1\neq\k_2\neq k_3$, 
and the Higgs fields $H_q$, $q=u,d$, present in the theory 
are singlets with respect to $S^\prime_4$, but 
may carry non-zero weights $k_{H_q}$. 
The modular invariance implies that the 
quark mass matrix $M_q$ should be formed by 
modular forms which furnish the ${\bf \hat{3}}^\prime$
representation of $S^\prime_4$. 
These are: the weigh 3 and 5 modular forms 
 $Y^{3}_{\hat{3}'}(\tau)$ and $Y^{5}_{\hat{3}'}(\tau)$, 
two weight 7 and two weigh 9 modular forms, 
$Y^{7(9)}_{\hat{3}',1}(\tau)$, $Y^{7(9)}_{\hat{3}',2}(\tau)$,
the three weight 11 modular forms  $Y^{11}_{\hat{3}',i}$, $i=1,2,3$, 
etc. \cite{Novichkov:2020eep}. 
Each modular form in $M_q$ is accompanied by a
constant. Consider the ``minimal'' case of 
four modular forms present in $M_q$ 
\footnote{ Quark (charged lepton) mass matrices which 
include only three different modular forms in the considered case 
of doublets superfields forming a triplet representation of 
$S^\prime_4$ and the three quark (charged lepton) 
singlet superfields furnishing singlet representations of 
 $S^\prime_4$, have one zero eigenvalue and thus 
are not phenomenologically viable without some additional 
input.
}. 
Choosing these to be the weight 3, 5 and 9 modular forms 
leads to ${\rm Det}(M_q) = 0$ and thus to 
one massless quark, which is incompatible 
with the existing data on, and lattice calculations 
of, the light quark masses. In this case one needs 
additional operators to generate non-zero 
mass for the lightest quark 
(see, e.g., \cite{Petcov:2023vws}).
Thus, in the considered case 
the only minimal and phenomenologically viable 
possibility involves the modular forms of 
weights 3, 5 and 7 and four constants.
The quark mass matrix $M_q$ 
has the following form in the R-L convention:
%%%%%%%%%%%%%%%%%%%%%%%%%%
\begin{align}
&  M_q =v_q
\frac{1}{\sqrt{3}}\,
\begin{pmatrix}
\alpha'_q  & 0 & 0  \\
0 & \beta_q & 0 \\
0& 0 & \gamma_q \\
\end{pmatrix}
\begin{pmatrix}
\tilde Y_{\hat{3}';1}^{(7)} & \tilde Y_{\hat{3}';3}^{(7)} & \tilde Y_{\hat{3}'2}^{(7)} \\
Y_{\hat{3}';1}^{(5)} & Y_{\hat{3}';3}^{(5)} & Y_{\hat{3}';2}^{(5)} \\
Y^{(3)}_{\hat{3}';1} & Y^{(3)}_{\hat{3}';3} & Y^{(3)}_{\hat{3}';2} 
\end{pmatrix}\,,
\label{eq:Mq31}
\end{align}
%%%%%%%%%%%%%%%%%%%%%%%%%%%%%%
% 
where $v_q$ is the Higgs doublet VEV,
$\tilde Y_{\hat{3}';i}^{(7)} =
Y_{\hat{3}',2;i}^{(7)} + g_q Y_{\hat{3}',1;i}^{(7)}$,   
$g_q = \alpha_q/\alpha'_q$,
$\alpha_q$, $\alpha'_q$, $\beta_q$
and $\gamma_q$ are the four constants,
and $Y_{\hat{3}',2;i}^{(7)}$, $Y_{\hat{3}',1;i}^{(7)}$,
$Y_{\hat{3}';i}^{(5)}$ and  $Y^{(3)}_{\hat{3}';i}$, $i=1,2,3$,
are the three components of the 
triplet modular forms. The mass matrix in Eq. (\ref{eq:Mq31}) 
leads in the ``vicinity'' of the fixed point 
$\tau_{\rm T} = i\infty$ to the following hierarchy between the three 
quark masses $1 : \epsilon^2 : \epsilon^3$, 
where $\epsilon \cong {\rm exp}(- \pi {\rm Im}\tau/2) \ll 1 $ 
is a measure of the ``deviation'' of the VEV $\tau_v$ of $\tau$ 
from $i\infty$ \cite{Novichkov:2021evw}. In the ``bottom-up'' modular 
flavour models the value of ${\rm Im}\tau_v$ is obtained from fits of 
the relevant data and typically one finds ${\rm Im}\tau_v \sim 2.5-3.0$ 
(see, e.g.,  \cite{Novichkov:2021evw,Abe:2023qmr,Petcov:2023vws})
\footnote{
We note also that $\psi$ can be chosen to be the three lepton 
doublet superfields, $\psi = L = (L_1, L_2, L_3)$, and 
$\psi^c$ to be the three singlet charged lepton fields 
 $\psi = E^c = (E^c_1, E^c_2, E^c_3)$. 
In this case the mass matrix in Eq. (\ref{eq:Mq31}) 
will lead to the hierarchy  $1 : \epsilon^2 : \epsilon^3$ 
of the charged lepton masses.
}.

  If one uses the ``minimal''  Kähler potential,
%%%%%%%%%%%%%%%%%%%%%%%%%%%%%%%%%%%
\begin{equation}
K(\tau, \overline{\tau}, \psi_I, \overline{\psi_I}) = 
-\, \Lambda_0^2\,\log(-i\tau + 
i\overline{\tau}) 
+ \sum_{I} \frac{|\psi_I|^2}{(-i\tau + i\overline{\tau})^{k_I}}\,,
\label{eq:Kahler}
\end{equation}
%%%%%%%%%%%%%%%%%%%%%%%%%%%
%
$\Lambda_0$ being a constant of mass dimension one,  
which is typically done within the bottom-up approach 
in the modular flavour model building,  
after the modulus $\tau$ gets a VEV $\tau_v$, the fermion fields 
have to be renormalised as follows in order to have a canonical 
kinetic term:
%%%%%%%%%%%%%%%%%%%%%%%
\begin{equation} 
\psi_I \rightarrow \sqrt{(2Im(\tau_v))^{k_I}} \psi_I\,.
\label{eq:psirenorm}
\end{equation}
%%%%%%%%%%%%%%%%%%%%%%%%%%%
%
Correspondingly, the bilinear term in eq. (\ref{eq:bilinear}) 
changes as:
%%%%%%%%%%%%%%%%%%%%%%
\begin{equation}
 \psi^c \, Y(\tau_v)\, \psi \rightarrow 
\sqrt{(2Im(\tau_v))^{k_\psi+ k^c_\psi}}\,\psi^c \, Y(\tau_v)\, \psi
\label{eq:renormbil}
\end{equation}
%%%%%%%%%%%%%%%%%%%%%%%%%%
%
This implies that in the considered example of a quark mass matrix  
in Eq. (\ref{eq:Mq31}), each constant gets an additional factor:
%%%%%%%%%%%%%%%%%%%%%%%%%%%%%%%%%%
\begin{eqnarray}
&&\alpha_q \rightarrow \hat\alpha_q= 
\alpha_q\, \sqrt{(2 {\rm Im} \tau_v)^{7}}\,,\\
&&\alpha'_q \rightarrow \hat\alpha'_q = 
\alpha'_q\, \sqrt{(2 {\rm Im} \tau_v)^{7} }\,,\\
&&\beta_q  \rightarrow \hat\beta_q = 
\beta_q  \, \sqrt{(2 {\rm Im} \tau_v)^{5}}\,,\\
&&\gamma_q  \rightarrow \hat\gamma_q = 
\gamma_q \sqrt{(2 {\rm Im} \tau_v)^{3}}\,.
\end{eqnarray}
%%%%%%%%%%%%%%%%%%%%%%%5
% 

We note that since in certain cases of models,  
in which the fermion (charged lepton and quark) mass 
hierarchies are obtained without fine tuning 
of the constants present in the fermion mass matrices
using the formalism developed in \cite{Novichkov:2021evw}
(see also \cite{Okada:2020ukr,Feruglio:2021dte}),  
$2{\rm Im} \tau_v$ can have a relatively large value. 
In the cases of models constructed in the 
``vicinity'' of the fixed point $\tau_{\rm T} = i\infty$ 
(see \cite{Novichkov:2018ovf,Novichkov:2018yse}),
for example, $2{\rm Im} \tau_v \sim 5$, as we have already indicated
(see, e.g., \cite{Novichkov:2021evw,Abe:2023qmr,Petcov:2023vws}).
This can lead to  significant modifications of the 
coupling constants: if initially they  
were of the same order in magnitude, 
they can become hierarchical, which in turn 
was used, e.g., in \cite{Abe:2023qmr,Petcov:2023vws}, to 
help explain the up-type quark mass hierarchies
in quark flavour models containing one modulus 
(for problems in constructing viable non-fine-tuned quark flavour 
models see, e.g., \cite{Petcov:2022fjf}).

  As we have already remarked, 
it is well known that the modular forms 
furnishing irreducible representations of the 
finite modular groups are determined up to a constant
In addition, since for a given level N, 
the higher weight modular forms are obtained as tensor products 
of the lowest weight modular form -- they are homogeneous 
polynomials of the lowest weight modular form -- 
the normalisation constant of the lowest weight modular 
form propagates in the higher weight modular forms. 
Usually, as we have indicated, these normalisation constants 
are absorbed in the constant parameters 
which multiply each modular form 
present in the fermion mass matrices. 
Since the normalisation of the modular forms 
is arbitrary, this makes the specific values of the 
constant parameters, obtained in a given model 
by statistical analysis of the
description of the relevant experimental data 
by the model, physically irrelevant.

Another problems is that some of the modular forms used 
in the ``non-fine-tuned models'' of flavour 
which are studied in the vicinity of one of the three 
known fixed points of the modular group 
\cite{Novichkov:2018ovf,Novichkov:2018yse}, 
namely, $\tau_C = i$, $\tau_L = \omega\equiv {\rm exp}(i2\pi/3)$ 
(the left cusp) and $\tau_T = i\infty$,  
vanish at the corresponding fixed point. 
In the case of models based on  $S^\prime_4$ and 
 $\tau = i\infty$, for example, 
we have \cite{Novichkov:2020eep,Novichkov:2021evw}
$Y_{\hat{1}'}^{(3)} = 0$, $Y_{3}^{(4)} = 0$, 
 $Y_{\hat{3},2}^{(5)} = 0$, $Y_{1'}^{(6)} = 0$,
$Y_{3}^{(6)} = 0$, $Y_{3',2}^{(6)} = 0$, 
 $Y_{\hat{1}'}^{(7)} = 0$, $Y_{\hat{2}}^{(7)} = 0$,
$Y_{\hat{3},2}^{(7)} = 0$, $Y_{\hat{3}',2}^{(7)} = 0$, 
etc. However, these modular forms can be 
normalised in the vicinity of the fixed point
(as well as at the fixed point) 
in such a way that with the new normalisation 
the singlet modular forms have a value of order 1,
$|\tilde{Y}_{\hat{1}'}^{(3)}| \sim 1$, 
$|\tilde{Y}_{1'}^{(6)}| \sim 1$, 
$|\tilde{Y}_{\hat{1}'}^{(7)}| \sim 1$,
while the triplet modular forms listed above 
have at least one component  
whose value becomes much larger 
(e.g., by one or two orders of magnitude, see further)
or even of order 1. 
If any of these modular forms plays an important role 
in making a given model phenomenologically viable, 
it should be obvious that this conclusion 
would depend on the normalisation 
employed of the modular form.
For this reason, in our opinion, modular forms 
which by changing their normalisation can change 
significantly their values 
(e.g., from a value close to zero or zero to a value $\sim 1$ ), 
which in turn can have phenomenological consequences, 
should be avoided in constructing non-fine-tuned models of flavour.

 In order to have a sensible comparison 
of the values of the constants obtained 
in different models when they are confronted with the data, 
we need to adopt a certain convention about the normalisation
of the modular forms. 

%%%%%%%%%%%%%%%%%%%%%%%%%%
%
\section{Modular Invariant Normalisations}
%
%%%%%%%%%%%%%%%%%%%%%%%%%%
%%%%%%%%%%%%%%%%%%%%%%%%%%
%
\subsection{``Local'' Normalisation at $\tau = \tau_v$}
%
%%%%%%%%%%%%%%%%%%%%%%%%%%%%

  Suppose we normalise the modular 
form of weight $K$, $Y^{(K)}(\tau)$,  by a factor $N^{(K)}_Y$, where 
%%%%%%%%%%%%%%%%%%%%%%
\begin{equation}
 N^{(K)}_Y 
= \left ( \sum_i |Y^{(K)}_i(\tau)|^2 (2Im(\tau))^{K} \right)^{\frac{1}{2}}\,,
\label{eq:Ny}
\end{equation}
%%%%%%%%%%%%%%%%%%%%%%%%
%
where $Y^{(K)}_i$ are the components of $Y^{(K)}$. Assuming that 
$Y^{(K)}$ furnishes a unitary representation {\bf r}
of the finite modular group of level N, 
 $Y^{(K)} = Y^{(K)}_{{\bf r}}$, 
this normalisation is modular invariant. 
Indeed, under the modular transformation,
%%%%%%%%%%%%%%%%%%%%%%%%%%%%
\begin{equation}
\begin{aligned}
\sum_i |Y^{(K)}_{{\bf r}i}|^2 \rightarrow \sum_i |Y^{(K)}_{{\bf r}i}|^2 
|(c\tau + d)|^{2K}\,,\\[0.25cm] 
(2Im(\tau))^{K} \rightarrow (2Im(\tau))^{K} |(c\tau + d)|^{-2K}\,.
\end{aligned}
\label{eq:NYmodinv}
\end{equation}
%%%%%%%%%%%%%%%%%%%%%%%%%%%%%
%

 In a specific flavour model, 
the modulus $\tau$ in the expression of $N^{(K)}_Y$ 
should be replaced by its VEV $\tau_v$ in the model, 
so the normalisation should read 
 $N^{(K)}_Y(\tau_v)$.
The quantity $N^{(K)}_Y$ determined in Eq. (\ref{eq:Ny})
seems an ``appropriate''  normalisation of 
$Y^{(K)}_{{\bf r}}(\tau)$. 
Using $N^{(K)}_Y(\tau_v))$ as a normalisation, has 
an additional important consequence.
Indeed, employing the described normalisation of $Y^{(K)}_{{\bf r}}(\tau)$
and taking into account the normalisation factor coming from the  
Kähler potential, we get for the bilinear of interest:
%%%%%%%%%%%%%%%%%%%%%%
\begin{equation}
 \psi^c \, Y^{(K)}_{{\bf r}}(\tau)\, \psi \rightarrow 
\sqrt{(2Im(\tau_v))^{k_\psi+ k^c_\psi}}\,\psi^c \, Y^{(K)}_{{\bf r}}(\tau_v)\, 
\psi/N^{(K)}_Y(\tau_v)
= \psi^c \, R_{\rm am}\,\frac{Y^{(K)}_{{\bf r}}(\tau)}
{\sqrt{\sum_i |Y^{(K)}_{{\bf}i}|^2}}\,\psi\,, 
\label{eq:renormbil2}
\end{equation}
%%%%%%%%%%%%%%%%%%%%%%%%%%
%
where $R_{\rm am}$ is the ratio of the automorphy factors 
due to the re-normalisation of $\psi$ and $\psi^c$ 
and the normalisation  (\ref{eq:Ny}) 
of the modular form:
%%%%%%%%%%%%%%%%%%%%%%%%%%%%
\begin{equation}
R_{\rm am} = \dfrac{\sqrt{(2Im(\tau_v))^{k_\psi+ k^c_\psi}}}
{\sqrt{(2Im(\tau_v))^K}}  = \dfrac{1}{\sqrt{(2Im(\tau_v))^{k_H}}}\,.
\label{eq:automr}
\end{equation}
%%%%%%%%%%%%%%%%%%%%%%%%%%%%%%%%
%
We see that the automorphy factors coming from 
the re-normalisation of $\psi$ and $\psi^c$
cancel in the ratio $R_{\rm am}$ 
as a consequence of $K = k_\psi + k^c_\psi + k_H$, 
which follows from the modular invariance.
This procedure (and cancellation) holds only after the modulus 
$\tau$ takes a VEV. If $k_H = 0$, as is assumed typically 
(but not universally, see, e.g., \cite{Feruglio:2021dte,Petcov:2023vws})
in bottom-up modular flavour models, we have  $R_{\rm am} = 1$.
We will consider this case in what follows.

 Under the assumed conditions, normalising
the modular forms to their Euclidean  norms at $\tau = \tau_v$
seems to be a ``natural'' normalisation. 
In that case the re-normalisation of the fermion fields 
in the Yukawa couplings, associated with obtaining 
their canonical kinetic terms, is effectively taken 
into account and does not need to be introduced in the 
fermion mass matrices. 
Moreover, one can use flavour-dependent weights 
of the matter fields in the non-fine-tuned approach to the fermion 
mass hierarchies \cite{Novichkov:2021evw}.
The Euclidean norms of the modular forms also 
do not depend on the basis one uses for the generators of 
the respective finite modular group. 

Let us consider a few examples of  $N^{(K)}_Y(\tau_v)$ 
in the case of level $N=4$ modular forms. 
The modular forms of interest are expressed  in terms of two 
holomorphic functions of $\tau$,
$\theta(\tau)$ and $\varepsilon(\tau)$ 
\cite{Novichkov:2020eep}:
%%%%%%%%%%%%%%%%%%%%%%%%%%%%
\begin{equation}
  \theta(\tau)\, \equiv\,
  \frac{\eta^5(2\tau)}
{\eta^2(\tau) \eta^2(4\tau)} \,=\, 
\Theta_3(2\tau) \,, \quad\,\,
\varepsilon(\tau) \,\equiv\, \frac{2\,\eta^2(4\tau)}{\eta(2\tau)} \,
=\, \Theta_2(2\tau) \,.
  \label{eq:theta_eps_def}
\end{equation}
%%%%%%%%%%%%%%%%%%%%%%%%%%%%%%%
%
Here $\eta(\tau)$ is the  Dedekind eta-function and 
$\Theta_2(\tau)$ and $\Theta_3(\tau)$ are the Jacobi theta constants.
The functions $\theta(\tau)$ and $\varepsilon(\tau)$ admit the following
$q$-expansions, i.e.~power series expansions in 
$q_4 \equiv \exp(i\pi \tau/2)$ 
\footnote{See Ref. \cite{Novichkov:2020eep} 
for technical details related to the properties 
of the functions  $\theta(\tau)$ and $\varepsilon(\tau)$.
}:
%%%%%%%%%%%%%%%%%%%%%%%%%%%
\begin{equation}
  \begin{aligned}
    \theta(\tau) \,&=\, 1
    + 2\sum_{k=1}^{\infty} q_4^{(2k)^2}
    \,=\, 1 + 2 \,q_4^4 + 2 \,q_4^{16} + \ldots \,, \\
    \varepsilon(\tau) \,&=\, 2\sum_{k=1}^{\infty} q_4^{(2k-1)^2} \,=\, 2 \,q_4^{\phantom{1}} + 2 \,q_4^9 + 2 \,q_4^{25} + \ldots \,,
  \end{aligned}
  \label{eq:theta_eps_qexp}
\end{equation}
%%%%%%%%%%%%%%%%%%%%%%%%%%%%%%
%
so that $\theta \to 1$, $\varepsilon \to 0$ in the ``large volume''
limit $\im\tau \to \infty$.
We give below the values of  $\theta(\tau)$
and $\varepsilon(\tau)$ at the three fixed point values of $\tau$,
namely, $\tau_C \equiv i$, 
$\tau_L \equiv \omega = {\rm exp}(i2\pi/3) = -\,1/2 + i\sqrt{3}/2$, 
and $\tau_T \equiv i\infty$, 
at which there exist residual symmetries 
used in the construction of non-fine-tuned modular flavour 
models: 
%%%%%%%%%%%%%%%%%%%%%%%%%%%%%%%%
\begin{equation}
\begin{aligned}
\theta(\tau_C) &= 1 + 2\,e^{-2\pi} + O(10^{-11}) \,\simeq\, 1.00373\,,\\
\varepsilon(\tau_C) &= 2\,e^{-\pi/2} + O(10^{-6})  \,\simeq\, 0.415761\,; 
% \\[3mm]
\end{aligned}
\nonumber
%\label{eq:thepsres0}
\end{equation}
%%%%%%%%%%%%%%%%%%%%%%%%
%%%%%%%%%%%%%%%%%%%%%%%%
\begin{equation}
\begin{aligned}
\theta(\tau_L) &=  1 - 2\,e^{-\sqrt{3}\, \pi}+ O(10^{-9}) \,\simeq\, 0.991333\,,\\
\varepsilon(\tau_L) &= 2\,e^{-i\,\pi/4}\left[e^{-\sqrt{3}\,\pi/4} + O(10^{-5})\right] \,\simeq\, 0.512152\;e^{-i\,\pi/4}\,;\\[3mm]
\theta(\tau_T) &= 1\,,\\\varepsilon(\tau_T) &= 0\,.
\end{aligned}
\label{eq:thepsres}
\end{equation}
%%%%%%%%%%%%%%%%%%%%%%%%%%%%%%%
%
We further give the exact relations between 
$\theta(\tau)$ and $\varepsilon(\tau)$ at the symmetric points 
$\theta(\tau_C)$ and $\tau_L$  :
%%%%%%%%%%%%%%%%%%%%%
\begin{equation}
    \frac{\varepsilon(\tau_C)}{\theta(\tau_C)} \,=\, \frac{1}{1+\sqrt{2}}\,,
    \qquad
    \frac{\varepsilon(\tau_L)}{\theta(\tau_L)} \,=\, \frac{1-i}{1+\sqrt{3}}\,.
\end{equation}
%%%%%%%%%%%%%%%%%%%%%%%%%%%%
%

The level $N=4$ lowest weight 1 modular form furnishes a 
${\bf \hat{3}}$ representation of $S^\prime_4$ \cite{Novichkov:2020eep}.
In the  group representation basis used in \cite{Novichkov:2020eep}  
this modular form is given by:
%%%%%%%%%%%%%%%%%%%%%%%%%
\begin{equation}
  Y_{\mathbf{\hat{3}}}^{(1)}(\tau) =
  \begin{pmatrix}
    \sqrt{2} \, \varepsilon \, \theta \\[1mm]
    \varepsilon^2 \\[1mm]
    -\theta^2
  \end{pmatrix}\,.
  \label{eq:k1triplet}
\end{equation}
%%%%%%%%%%%%%%%%%%%%%%%%%%%%
%
Further tensor products with $Y_{\mathbf{\hat{3}}}^{(1)}$ produce new modular
multiplets of even and odd weights
\footnote{In what follows we use the expressions for the modular 
form multiplets derived in \cite{Novichkov:2020eep}, 
where the multiplets with weights $k_Y\leq 10$ have been constructed.
}.
We give below the expressions of selected level 4 weight 3, 4, 5 and 7 
modular forms which we use for illustrative purposes 
in our discussion.  
At weight $k=3$ there are three multiplets - a non-trivial singlet and two
triplets - exclusive to $S_4'$:
%%%%%%%%%%%%%%%%%%%%%%%%%%%%%%%%
\begin{equation}
\begin{aligned}
  Y_{\mathbf{\hat{1}'}}^{(3)}(\tau) &= \sqrt{3} \left(
  \varepsilon \,\theta^5-\varepsilon^5\, \theta
  \right)\,,
  \\[2mm]
  Y_{\mathbf{\hat{3}}}^{(3)}(\tau) &=
  \begin{pmatrix}
 \varepsilon ^5\, \theta +\varepsilon \, \theta ^5\\[1mm]
 \frac{1}{2\sqrt{2}}\left(5 \,\varepsilon ^2 \, \theta ^4-\varepsilon ^6 \right)\\[1mm]
 \frac{1}{2\sqrt{2}}\left(\theta ^6-5\, \varepsilon ^4 \,\theta ^2\right)
  \end{pmatrix}\,,\quad
    Y_{\mathbf{\hat{3}'}}^{(3)}(\tau) = \frac{1}{2}
  \begin{pmatrix}
 -4 \sqrt{2}\, \varepsilon ^3 \,\theta ^3 \\[1mm]
 \theta ^6 + 3 \,\varepsilon ^4\, \theta ^2 \\[1mm]
 - 3\, \varepsilon ^2\, \theta ^4 -\varepsilon ^6
  \end{pmatrix}\,.
\end{aligned}
\label{eq:k3}
\end{equation}
%%%%%%%%%%%%%%%%%%%%%%%%%%%%%%%%%%%%
% 
The set of four weight $k=4$ modular forms includes:
 \cite{Penedo:2018nmg,Novichkov:2020eep}:
%%%%%%%%%%%%%%%%%%%%%%%%%%%%%%%%%%%
\begin{equation}
\begin{aligned}
 % Y_{\mathbf{1}}^{(4)}(\tau) &=
 % \frac{1}{2 \sqrt{3}} \left(
 % \theta^8 + 14\, \varepsilon^4\, \theta^4 + \varepsilon^8
 % \right)\,,\quad
 % Y_{\mathbf{2}}^{(4)}(\tau) =
 % \begin{pmatrix}
 % \frac{1}{4} \left(\theta^8 - 10 \,\varepsilon^4\, \theta^4 + \varepsilon^8\right) \\[1mm]
 % \sqrt{3}\left(\varepsilon ^2\,\theta^6 + \varepsilon ^6\, \theta^2\right)
 %  \end{pmatrix}\,,
 % \\[2mm]
  Y_{\mathbf{3}}^{(4)}(\tau) &=
  \frac{3}{2\sqrt{2}}
  \begin{pmatrix}
 \sqrt{2}\left(\varepsilon ^2\, \theta^6 -\varepsilon^6 \, \theta ^2\right)\\[1mm]
 \varepsilon ^3 \,\theta ^5 - \varepsilon ^7 \,\theta \\[1mm]
 - \varepsilon  \,\theta ^7 + \varepsilon ^5 \,\theta ^3
  \end{pmatrix}\,,\quad
  Y_{\mathbf{3'}}^{(4)}(\tau) =
  \begin{pmatrix}
  \frac{1}{4}\left(\theta ^8-\varepsilon ^8\right)\\[1mm]
  \frac{1}{2\sqrt{2}}\left(\varepsilon \, \theta ^7 + 7 \,\varepsilon ^5\, \theta ^3\right) \\[1mm]
  \frac{1}{2\sqrt{2}}\left(7 \,\varepsilon ^3\, \theta ^5 + \varepsilon ^7 \,\theta\right)
  \end{pmatrix}\,,
\end{aligned}
\label{eq:k4}
\end{equation}
%%%%%%%%%%%%%%%%%%%%%%%%%%%%%%%
%

The weight $k=5$ forms of interest are:
%%%%%%%%%%%%%%%%%%%%%%%%%%%%%%%%%
\begin{equation}
\begin{aligned}
  Y_{\mathbf{\hat{2}}}^{(5)}(\tau) &=
  \begin{pmatrix}
  \frac{3}{2}\left( \varepsilon ^3 \,\theta ^7-\varepsilon ^7 \,\theta ^3 \right)\\[1mm]
  \frac{\sqrt{3}}{4}\left( \varepsilon  \,\theta ^9-\varepsilon ^9 \,\theta   \right)
  \end{pmatrix}\,,
 \\[2mm]
   Y_{\mathbf{\hat{3}},1}^{(5)}(\tau) &=
  \begin{pmatrix}
  \frac{6\sqrt{2}}{\sqrt{5}} \,
 \varepsilon ^5\, \theta ^5
 \\[1mm]
 \:\:\:\frac{3}{8\sqrt{5}} \left(
 5\, \varepsilon ^2\, \theta ^8
 +10\, \varepsilon ^6\, \theta ^4
 +\varepsilon ^{10}
 \right) \\[1mm]
 -\frac{3}{8\sqrt{5}} \left(
 \theta ^{10}
 +10\, \varepsilon ^4\, \theta ^6
 +5\, \varepsilon ^8\, \theta ^2
 \right)
  \end{pmatrix}\,,
% \\[2mm]
\end{aligned}
% \label{eq:k50}
\nonumber
\end{equation}
%%%%%%%%%%%%%%%%%%%%%%%%%
%%%%%%%%%%%%%%%%%%%%%%%%%
\begin{equation}
\begin{aligned}
%  \\[2mm]
  Y_{\mathbf{\hat{3}},2}^{(5)}(\tau) &=
  \begin{pmatrix}
  \frac{3}{4}\left( \varepsilon  \, \theta ^9 -2 \,\varepsilon ^5 \, \theta ^5 + \varepsilon ^9 \, \theta \right)\\[1mm]
  \frac{3}{\sqrt{2}}\left(-\varepsilon ^2 \, \theta ^8 + \varepsilon ^6 \, \theta ^4\right)\\[1mm]
  \frac{3}{\sqrt{2}}\left(-\varepsilon ^4 \, \theta ^6 + \varepsilon ^8 \, \theta ^2\right)
  \end{pmatrix}\,,
  \\[2mm]
  Y_{\mathbf{\hat{3}'}}^{(5)}(\tau) &=
  \begin{pmatrix}
  2\left(\varepsilon ^3\, \theta ^7 +  \varepsilon ^7 \,\theta ^3\right)\\[1mm]
  \frac{1}{4\sqrt{2}}\left(\theta ^{10} -14 \,\varepsilon ^4\, \theta ^6  -3 \,\varepsilon ^8\, \theta ^2 \right) \\[1mm]
  \frac{1}{4\sqrt{2}}\left( 3\, \varepsilon ^2\, \theta ^8 + 14\, \varepsilon ^6\, \theta ^4 -\varepsilon ^{10 }\right)
  \end{pmatrix}\,.
\end{aligned}
\label{eq:k5}
\end{equation}
%%%%%%%%%%%%%%%%%%%%%%%%%%%%%
%
In the case of weight $k=7$ we have:
%%%%%%%%%%%%%%%%%%%%%%%%%%%%%%%%%%
\begin{equation}
\begin{aligned}
  Y_{\mathbf{\hat{1}'}}^{(7)}(\tau) &=
  \frac{1}{4} \sqrt{\frac{3}{2}} \left(
-\varepsilon ^{13}\, \theta
-13\, \varepsilon ^9\, \theta ^5
+13\, \varepsilon ^5\, \theta ^9
+\varepsilon\,  \theta ^{13}
\right)
  \,,\\[2mm]
  Y_{\mathbf{\hat{2}}}^{(7)}(\tau) &=
  \begin{pmatrix}\frac{3}{2} \left(
\varepsilon ^3\, \theta ^{11}
-\varepsilon ^{11} \,\theta ^3
\right) \\[1mm]
-\frac{\sqrt{3}}{8} \left(
\varepsilon\,  \theta ^{13}
-11\, \varepsilon ^5\, \theta ^9
+11\, \varepsilon ^9\, \theta ^5
-\varepsilon ^{13} \,\theta
\right)
  \end{pmatrix}\,,
  \\[2mm]
  Y_{\mathbf{\hat{3}},1}^{(7)}(\tau) &=
  \begin{pmatrix}
  \frac{12}{\sqrt{13}} \left(
\varepsilon ^5\, \theta ^9
 +\varepsilon ^9\, \theta ^5
\right) \\[1mm]
\frac{3}{8\sqrt{26}} \left(
\varepsilon ^2\, \theta ^{12}
 +45\, \varepsilon ^6\, \theta ^8
 +19\, \varepsilon ^{10} \,\theta ^4
 -\varepsilon ^{14}
\right) \\[1mm]
\frac{3}{8\sqrt{26}} \left(
 \theta ^{14}
 -19\, \varepsilon ^4\, \theta ^{10}
 -45\, \varepsilon ^8\, \theta ^6
 -\varepsilon ^{12}\, \theta ^2
\right)
\end{pmatrix}\,,
  \\[2mm]
  Y_{\mathbf{\hat{3}},2}^{(7)}(\tau) &=
  \begin{pmatrix}
  \frac{3}{8} \left(
 \varepsilon\,  \theta ^{13}
 -\varepsilon ^5\, \theta ^9
 -\varepsilon ^9\, \theta ^5
 +\varepsilon ^{13}\, \theta
\right) \\[1mm]
\frac{3}{4\sqrt{2}} \left(
 \varepsilon ^2\, \theta ^{12}
 +6\, \varepsilon ^6\, \theta ^8
 -7\, \varepsilon ^{10} \,\theta ^4
\right) \\[1mm]
\frac{3}{4\sqrt{2}} \left(
 7\, \varepsilon ^4\, \theta ^{10}
 -6\, \varepsilon ^8\, \theta ^6
 -\varepsilon ^{12} \,\theta ^2
\right)
  \end{pmatrix}\,,
  \\[2mm]
  Y_{\mathbf{\hat{3}'},1}^{(7)}(\tau) &=
  \begin{pmatrix}
  \frac{3}{4\sqrt{37}}\left(
  7\, \varepsilon ^3 \,\theta ^{11}
 +50\, \varepsilon ^7\, \theta ^7
 +7\, \varepsilon ^{11}\, \theta ^3
\right) \\[1mm]
-\frac{3}{4\sqrt{74}} \left(
\theta ^{14}
 +14\, \varepsilon ^4\, \theta ^{10}
 +49\, \varepsilon ^8\, \theta ^6
\right) \\[1mm]
\:\:\:\frac{3}{4\sqrt{74}} \left(
49\, \varepsilon ^6\, \theta ^8
 +14\, \varepsilon ^{10} \,\theta ^4
 +\varepsilon ^{14}
\right)
\end{pmatrix}\,,
  \\[2mm]
  Y_{\mathbf{\hat{3}'},2}^{(7)}(\tau) &=
  \begin{pmatrix}
\frac{9}{4}
\left(  \varepsilon ^3\, \theta ^{11}
 -2\, \varepsilon ^7\, \theta ^7
 +\varepsilon ^{11} \,\theta ^3 \right)
\\[1mm]
\:\:\:\frac{9}{4\sqrt{2}} \left(
\varepsilon ^4\, \theta ^{10}
 -2\, \varepsilon ^8\, \theta ^6
 +\varepsilon ^{12}\, \theta ^2
\right) \\[1mm]
-\frac{9}{4\sqrt{2}} \left(
\varepsilon ^2\, \theta ^{12}
 -2\, \varepsilon ^6\, \theta ^8
 +\varepsilon ^{10} \,\theta ^4
\right)
  \end{pmatrix}\,.
\end{aligned}
\label{eq:k7}
\end{equation}
%%%%%%%%%%%%%%%%%%%%%%%%%%%%%%%%%%%%%%%%
%

 In the expressions for the modular forms in 
Eqs. (\ref{eq:k3}) - (\ref{eq:k7})                
we have not included the overall arbitrary 
constant factors, which we assume to be real.
They will be present in the respective modular form 
normalisation constants and therefore will be canceled 
in the expressions for the normalised modular forms. 
Thus, the normalised modular forms will not depend 
on these arbitrary constants. 

Consider the effect of the normalisation 
by $N^{(K)}_Y(\tau_v)$ of level $N=4$  
modular forms in the vicinity 
of the fixed point $\tau_T = i\infty$.
For concreteness we fix ${\rm Im}\tau_v = 2.5$, 
as suggested by flavour models considered in the 
literature (see, e.g., 
\cite{Novichkov:2021evw,Abe:2023qmr,Petcov:2023vws})),   
while ${\rm Re}\tau_v$ is left arbitrary.
In this case 
$|q_{4v}| = {\rm exp(-\pi Im\tau_v/2)} \cong 0.019703$, 
and $|q_{4v}|^4 \cong 1.507\times 10^{-7}$.   
Therefore, to an excellent approximation 
$\theta(\tau_v) \cong 1$ and 
$\varepsilon(\tau_v) \cong 2q_{4v} 
= 2\,e^{i\,\pi {\rm Re}\tau_v/2}\,e^{-\,\pi {\rm Im}\tau_v/2}$.
In the considered case, for the normalisation factor 
$N^{(K)}_Y(\tau_v)$ of the modular forms 
$Y_{\mathbf{\hat{3}}}^{(1)}(\tau)$,
$Y_{\mathbf{\hat{1}'}}^{(3)}(\tau)$, 
$Y_{\mathbf{3}}^{(4)}(\tau)$,
$Y_{\mathbf{\hat{3}},2}^{(5)}(\tau)$ 
and 
$Y_{\mathbf{\hat{3}'},2}^{(7)}(\tau)$, for example, 
we get:
%%%%%%%%%%%%%%%%%%%%%%%%%%%%%%%%%%%
\begin{equation}
\begin{aligned}
& |Y_{\mathbf{\hat{3}}}^{(1)}(\tau_v)| \cong 
|\theta(\tau_v)|^2 (1 + 2|\varepsilon(\tau_v)|^2/|\theta(\tau_v)|^2)^{\frac{1}{2}} 
% \cong 1 + 7.76\times 10^{-4} 
= 1 + O(10^{-3})\,,
\\
& |Y_{\mathbf{\hat{1}'}}^{(3)}(\tau_v)| \cong 
\sqrt{3}|\varepsilon(\tau_v)||\theta(\tau_v)|^5\,,
\\
& |Y_{\mathbf{3}}^{(4)}(\tau_v)| \cong 
\frac{3}{2\sqrt{2}}\,|\varepsilon(\tau_v)||\theta(\tau_v)|^7\,,
\\
& |Y_{\mathbf{\hat{3}},2}^{(5)}(\tau_v)| \cong 
\frac{3}{4}\,|\varepsilon(\tau_v)||\theta(\tau_v)|^9\,, 
\\
& |Y_{\mathbf{\hat{3}'},2}^{(7)}(\tau_v)| \cong 
\frac{9}{4\sqrt{2}}|\varepsilon(\tau_v)|^2|\theta(\tau_v)|^{12}\,,
\end{aligned}
\end{equation}
%%%%%%%%%%%%%%%%%%%%%%%%%%%%%%
%
where the notation used for $N^{(K)}_Y(\tau_v)$ is 
self explanatory.

 Since in the considered case the normalisation of the lowest 
weight modular form 
$|Y_{\mathbf{\hat{3}}}^{(1)}(\tau_v)| = 1 +  O(10^{-4})$,
the effect of propagation of this normalisation to higher weight 
modular form is practically negligible.
The normalised $Y_{\mathbf{\hat{3}}}^{(1)}(\tau_v)$ has the form:
%%%%%%%%%%%%%%%%%%%%%%%%%%%%%%%%%%%%%
\begin{equation}
\dfrac{Y_{\mathbf{\hat{3}}}^{(1)}(\tau_v)}
{|Y_{\mathbf{\hat{3}}}^{(1)}(\tau_v)|}
=
 \begin{pmatrix}
    \sqrt{2}\, \frac{\varepsilon_v\,\theta_v}{|\theta_v|^2} \\[1mm]
  \frac{\varepsilon^2_v}{|\theta_v|^2} \\[1mm]
    -\,\frac{\theta^2_v}{|\theta_v|^2}
  \end{pmatrix}
\cong
\begin{pmatrix}
    \sqrt{2}\,\varepsilon_v \\[1mm]
  \varepsilon^2_v \\[1mm]
    -\,1
  \end{pmatrix}\,,
\label{eq:k1tripletNorm}
\end{equation}
%%%%%%%%%%%%%%%%%%%%%%%%%%%%
%
where we have used $\theta_v \cong 1$
(see Eq. (\ref{eq:theta_eps_qexp}) and the discussion 
after Eq. (\ref{eq:k7})).

If we normalise the singlet modular form $Y_{\mathbf{\hat{1}'}}^{(3)}(\tau)$ 
at $\tau = \tau_v$, we get:
%%%%%%%%%%%%%%%%%%%%%%%
\begin{equation}
 \dfrac{Y_{\mathbf{\hat{1}'}}^{(3)}(\tau_v)}{|Y_{\mathbf{\hat{1}'}}^{(3)}(\tau_v)|}
= \dfrac{\varepsilon_v \,\theta^5_v}{|\varepsilon_v \,\theta^5_v|}
 - \dfrac{\varepsilon^5_v}{|\varepsilon_v|}\,\dfrac{\theta_v}{|\theta^5_v|} 
\cong e^{i\,\pi\,Re(\tau_v)/2} \left ( 1 - \varepsilon^4_v)\right )\,.
\label{eq:Y1p3n}
\end{equation}
%%%%%%%%%%%%%%%%%%%%%%%%
%
We see that the used normalisation changes 
drastically the magnitude of this singlet modular form 
from being of order $|\varepsilon| \ll 1$ 
to being of order 1.

Consider next the effect of the normalisation on the 
triplet modular forms 
$Y_{\mathbf{3}}^{(4)}(\tau)$,
$Y_{\mathbf{\hat{3}},2}^{(5)}(\tau)$ and 
$Y_{\mathbf{\hat{3}'},2}^{(7)}(\tau)$:
%%%%%%%%%%%%%%%%%%%%%%%%%%%%%%%%%%%
\begin{equation}
\begin{aligned} 
\dfrac{Y_{\mathbf{3}}^{(4)}(\tau_v)}{|Y_{\mathbf{3}}^{(4)}(\tau_v)|} &=
\frac{1}{|\varepsilon(\tau_v)||\theta(\tau_v)|^7}
  \begin{pmatrix}
 \sqrt{2}\left(\varepsilon ^2_v\, \theta^6_v -\varepsilon^6_v \, \theta ^2\right)
\\[1mm]
 \varepsilon^3_v \,\theta^5_v - \varepsilon^7_v \,\theta \\[1mm]
 - \varepsilon_v\,\theta^7_v + \varepsilon^5_v\,\theta^3_v
  \end{pmatrix}
\cong
e^{i\,\pi {\rm Re}\tau/2}
\begin{pmatrix}
   \sqrt{2}\varepsilon_v     
\\[1mm]
 \varepsilon^2_v  
\\[1mm]
 -\,1 
  \end{pmatrix}\,,
\end{aligned}
\nonumber
\end{equation}
%%%%%%%%%%%%%%%%%%%%%%%%%%%%%%%%% 
%%%%%%%%%%%%%%%%%%%%%%%%%%%%%%%%%
\begin{equation}
\begin{aligned} 
% \\[2mm]
 \dfrac{Y_{\mathbf{\hat{3}},2}^{(5)}(\tau_v)}{|Y_{\mathbf{\hat{3}},2}^{(5)}(\tau_v)|} &=
\frac{4}{3\,|\varepsilon(\tau_v)||\theta(\tau_v)|^9}
  \begin{pmatrix}
  \frac{3}{4}\left( \varepsilon_v\,\theta^9_v -\,2\,\varepsilon^5_v \,
\theta^5_v + \varepsilon^9_v\,\theta_v \right)\\[1mm]
  \frac{3}{\sqrt{2}}\left(-\varepsilon^2_v \,\theta^8_v + 
\varepsilon^6_v \,\theta^4_v\right)\\[1mm]
  \frac{3}{\sqrt{2}}\left(-\varepsilon^4_v\,\theta^6_v 
+ \varepsilon^8_v\,\theta^2_v\right)
  \end{pmatrix}
\cong 
 e^{i\,\pi {\rm Re}\tau/2}  
\begin{pmatrix}
1  
\\[1mm]
 -\,\frac{4}{\sqrt{2}}\,\varepsilon_v 
\\[1mm]
-\,\frac{4}{\sqrt{2}}\,\varepsilon^3_v   
  \end{pmatrix}\,,
\\[2mm]
\dfrac{Y_{\mathbf{\hat{3}'},2}^{(7)}(\tau_v)}{|Y_{\mathbf{\hat{3}'},2}^{(7)}(\tau_v)|} &=
\frac{4\sqrt{2}}{9|\varepsilon(\tau_v)|^2|\theta(\tau_v)|^{12}}\,
\begin{pmatrix}
\frac{9}{4}
\left(\varepsilon^3_v\,\theta^{11}_v
 -\,2\,\varepsilon^7_v\, \theta^7_v
 +\varepsilon^{11}_v \,\theta^3_v \right)
\\[1mm]
\:\:\:\frac{9}{4\sqrt{2}} \left(
\varepsilon^4_v\,\theta^{10}_v
 -\,2\,\varepsilon^8_v\,\theta^6_v
 +\varepsilon^{12}_v\,\theta^2_v
\right) \\[1mm]
-\,\frac{9}{4\sqrt{2}} \left(
\varepsilon^2_v\,\theta^{12}_v
 -\,2\,\varepsilon^6_v\,\theta^8_v
 +\varepsilon^{10}_v\,\theta^4_v
\right)
  \end{pmatrix}
\cong 
 e^{i\,\pi {\rm Re}\tau}
\begin{pmatrix}
\sqrt{2}\,\varepsilon_v   
\\[1mm]
\:\:\:
  \varepsilon^2_v   
\\[1mm]
  -\,1 
 \end{pmatrix}\,.
\end{aligned}
\end{equation}
%%%%%%%%%%%%%%%%%%%%%%%%%%%%%%%%%%
%
We see that in the normalised weight 4 and 5 triplet modular forms
all components are increased by the factor 
$|\varepsilon(\tau_v)|^{-1} >> 1$, while in the weight 7 modular form the 
components are increased by $|\varepsilon(\tau_v)|^{-2} >> 1$.
As a consequence, instead of being of order $|\varepsilon(\tau_v)| \ll 1$ 
($|\varepsilon(\tau_v)|^2 \ll 1$),  
the 3rd and the 1st (the 3rd) components respectively of 
the weight 4 and 5 (weight 7) normalised triplet modular forms 
are of order 1. If any of these three triplet 
modular forms is present in a given fermion mass matrix, 
the results obtained by diagonalising the mass matrix, 
i.e., the mass eigenvalues and the values of the  
elements of the diagonalising unitary or orthogonal 
matrix, will depend on the chosen normalisation 
of this modular form. 

Indeed, consider our example of fermion mass matrix given in
Eq. (\ref{eq:Mq31}). It contains the modular form 
$Y_{\mathbf{\hat{3}'},2}^{(7)}(\tau)$ in addition to the modular forms 
$Y_{\mathbf{\hat{3}'},1}^{(7)}(\tau)$, $Y_{\mathbf{\hat{3}'}}^{(5)}(\tau)$ 
and $Y_{\mathbf{\hat{3}'}}^{(3)}(\tau)$.
If we use the expressions for these four modular forms 
given in Eqs. (\ref{eq:k3}),  (\ref{eq:k5}) and (\ref{eq:k3}), 
we find that in the limit of $|\varepsilon(\tau_v)|\rightarrow 0$ 
(${\rm Im}\tau_v \rightarrow \infty$), 
$Y_{\mathbf{\hat{3}'},2}^{(7)}(\tau_v) = 0$ and that  
only the second components of 
 $Y_{\mathbf{\hat{3}'},1}^{(7)}(\tau)$, $Y_{\mathbf{\hat{3}'}}^{(5)}(\tau)$ 
and $Y_{\mathbf{\hat{3}'}}^{(3)}(\tau)$ are non-zero and are of order 
$3/(4\sqrt{74})$, $1/(4\sqrt{2})$ and 1/2, respectively.
As a consequence, in the considered limit only the [3,3] element of 
the matrix $M^\dagger_qM_q$, whose diagonalising matrix contributes 
to the CKM quark mixing matrix, is non-zero. This implies that 
$M_q$ has only one ``large'' eigenvalue, $m_3$, which is given by
%%%%%%%%%%%%%%%%%%%%%%%%%%
\begin{equation}
m_3 = \dfrac{v_q}{4\sqrt{6}} 
\left ( \frac{9}{37}\, |\alpha_q|^2 + |\beta_q|^2 
+ 8\,|\gamma_q|^2\right)^{\frac{1}{2}}\,. 
\label{eq:m3}
\end{equation}
%%%%%%%%%%%%%%%%%%%%%%%
%
The other two eigenvalues, $m_{1,2}$, as can be shown, are by the 
factors $|\varepsilon(\tau_v)|^2$ 
and $|\varepsilon(\tau_v)|^3$ smaller \cite{Novichkov:2021evw}, which, 
in principle, may  allow to explain the hierarchy of quark
(or charged lepton) masses without fine-tuning of the constants 
$\alpha_q$, $\alpha'_q$, $\beta_q$ and $\gamma_q$ present in $M_q$.

 We get a completely different result if we use the expressions for 
the normalised modular forms  
$Y_{\mathbf{\hat{3}'}}^{(3)}(\tau_v)/|Y_{\mathbf{\hat{3}'}}^{(3)}(\tau_v)| $, 
$Y_{\mathbf{\hat{3}'}}^{(5)}(\tau_v)/|Y_{\mathbf{\hat{3}'}}^{(5)}(\tau_v)|$,
$Y_{\mathbf{\hat{3}'},1}^{(7)}(\tau_v)/|Y_{\mathbf{\hat{3}'},1}^{(7)}(\tau_v)|$ and
$Y_{\mathbf{\hat{3}'},2}^{(7)}(\tau_v)/|Y_{\mathbf{\hat{3}'},2}^{(7)}(\tau_v)|$ in $M_q$.
In the limit
$|\varepsilon(\tau_v)|\rightarrow 0$,  
again only the 2nd components of 
$Y_{\mathbf{\hat{3}'}}^{(3)}(\tau_v)/|Y_{\mathbf{\hat{3}'}}^{(3)}(\tau_v)| $, 
$Y_{\mathbf{\hat{3}'}}^{(5)}(\tau_v)/|Y_{\mathbf{\hat{3}'}}^{(5)}(\tau_v)|$,
$Y_{\mathbf{\hat{3}'},1}^{(7)}(\tau_v)/|Y_{\mathbf{\hat{3}'},1}^{(7)}(\tau_v)|$ 
are non-zero. Now they are  of order 
1, 1 and (-1), respectively. The third component 
of $Y_{\mathbf{\hat{3}'},2}^{(7)}(\tau_v)/|Y_{\mathbf{\hat{3}'},2}^{(7)}(\tau_v)|$
is also non-zero in this case 
and is given by $(-\,{\rm exp}(i\pi {\rm Re}\tau_v))$.
The matrix of interest $M^\dagger_qM_q$ can be cast in the form:
%%%%%%%%%%%%%%%%%%%%%%%%%%
\begin{align}
M^\dagger_qM_q 
=  v_q^2\, 
\begin{pmatrix}
0  & 0 & 0  \\
0 & |\alpha'_q|^2 & |\alpha_q\alpha'_q|\,e^{-\,i(\kappa_v + \kappa_g)}  \\
0&  |\alpha_q\alpha'_q|\,e^{i(\kappa_v + \kappa_g)} &  |\alpha_q|^2
+ |\beta_q|^2+|\gamma_q|^2 
\end{pmatrix}\,,
\label{eq:Mq2Norm}
\end{align}
%%%%%%%%%%%%%%%%%%%%%%%
%
where $\kappa_v = \pi {\rm Re}\tau_v$ and 
$\kappa_g = -\,{\rm arg}(g_q)  = -\,{\rm arg}(\alpha_q/\alpha'_q)$.
$M^\dagger_qM_q$ in Eq. (\ref{eq:Mq2Norm}) 
(and thus  $M_q$) has two ``large'' mass eigenvalues $m^2_{2,3}$ 
($m_{2,3}$).
The hierarchy $m_{2}/m_3 \ll 1$ can be obtained in the considered case
only by fine-tuning the values of the constants  
$|\alpha_q|$, $|\alpha'_q|$ and $|\beta_q|^2+|\gamma_q|^2$.
One possibility is, e.g.,  
$|\alpha_q| \cong |\alpha'_q| \ll (|\beta_q|^2+|\gamma_q|^2)^{\frac{1}{2}}$ 
\footnote{ In order for the (2-3) mixing
to be sufficiently small in the quark case, 
we should have also 
$|\alpha_q\alpha'_q| \ll (|\alpha_q|^2
+ |\beta_q|^2+|\gamma_q|^2 -  |\alpha'_q|^2)$.
}.
In this case to a good approximation 
$m_2/m_3 \cong (|\alpha_q|^2 + |\alpha'_q|^2)^{\frac{1}{2}}/
(\sqrt{2} (|\beta_q|^2+|\gamma_q|^2)^{\frac{1}{2}} )$.

%%%%%%%%%%%%%%%%%%%%%%%%%%
%
\subsection{``Global'' (or ``Integral'') Normalisation}
%
%%%%%%%%%%%%%%%%%%%%%%%%%%%%

 In the proposed ``local'' solution to the modular form 
normalisation problem 
within the non-fine-tuned approach to the fermon mass hierarchies in
the modular flavour models, the normalisation factors $N^{(K)}_Y$
depend on the VEV of the modulus $\tau$, 
$N^{(K)}_Y= N^{(K)}_Y(\tau_v)$.
If the model is constructed in the vicinity of a given 
fixed point, $N^{(K)}_Y(\tau_v)$ will depend on the chosen fixed point. 
That will allow meaningful comparison 
of different models constructed in the vicinity 
of the same fixed point, but somewhat ambiguous 
the comparison of models constructed at different 
fixed points. 

On the basis of the expression for $N^{(K)}_Y$, 
Eq. (\ref{eq:Ny}), one can form a modular invariant normalisation 
of the modular forms which does not depend 
on the VEV of the modulus $\tau$ and thus on 
the fixed point in the vicinity of which the model is constructed.
This can be done  by integrating 
$(N^{(K)}_Y)^2$
over the fundamental domain $D$ of the modular group using the 
modular invariant hyperbolic 
measure (or volume form) $d\mu(\tau) = dx\,dy/y^2$, 
where  $x\equiv {\rm Re} \tau$ and 
$y \equiv {\rm Im} \tau$:
%%%%%%%%%%%%%%%%%%%%%%%%%%%%%%%%
\begin{equation}
(\overline{{\rm N}}^{(K)}_{\rm Y })^2 
=  \int_{D}\int (N^{(K)}_{\rm Y} (\tau,\tau^*))^2\,\frac{dx dy}{y^2}
\equiv \int_{D} \int \sum_i |Y^{(K)}_{{\bf r}i}(\tau)|^2 (2\,y)^{K} \frac{dx\,dy}{y^2}\,.
\label{eq:IntNy2}
\end{equation}
%%%%%%%%%%%%%%%%%%%%%%%%%%%%%%%%%%%%%%
%
 The integration should be understood as integration over all possible 
VEVs of $\tau$ in the fundamental domain of the modular group
\footnote{A normalisation 
with integration over the fundamental domain ${\rm D}$ with the 
modular invariant hyperbolic 
measure $d\mu(\tau) = dx\,dy/y^2$ and integrand  $N^{(K)}_Y$ 
given in Eq. (\ref{eq:Ny}) was suggested to the author 
by P. Novichkov after the author proposed in July of 2022 and 
discussed with P. Novichkov and J. Penedo the normalisation 
defined in Eq. (\ref{eq:Ny}).
}.
The normalisation squared $(\overline{{\rm N}}^{(K)}_{\rm Y })^2$ 
is a particular case of what is known in mathematical literature  
as ``Petersson inner product of two modular forms'' 
\cite{Pet1932}
\footnote{The author learned this fact from Dr. A. M. Kidambi 
during the 2023 spring visit of Kavli IPMU, University of Tokyo.
}.
The integral in  Eq. (\ref{eq:IntNy2})
is independent of the choice of the fundamental domain
since both $\sum_i |Y_{{\bf r}i}(\tau)|^2 (2\,y)^{K}$ and 
the hyperbolic measure  $d\mu = dx\,dy/(y^2)$ are modular invariant.

 We show next that the hyperbolic measure $d\mu = dxdy/y^2$ is 
indeed modular invariant. Consider two moduli $z = u + i\,v$ and 
$\tau = x + iy$
related by a modular transformation:
%%%%%%%%%%%%%%%%%%%%%%%%%%
\begin{equation}
z = \frac{a\tau + b}{c\tau +d}\,,~~{\rm with~integer~a,b,c,d}\,~
{\rm and~ ad-bc = 1}\,.
\label{eq:ztau}
\end{equation}
%%%%%%%%%%%%%%%%%%%%%%%%%%
%
In the context of the problem we are considering 
Eq. (\ref{eq:ztau}) describes a change of variables 
$(u,v)$ to $(x,y)$:
%%%%%%%%%%%%%%%%%%%%%%%%%%
\begin{equation}
u = \frac{ac(x^2 + y^2) + x(ad + bc) + bd}{|c\tau +d|^2}\,,~~
v= \frac{y}{|c\tau +d|^2}\,.
\label{eq:uvxy}
\end{equation}
%%%%%%%%%%%%%%%%%%%%%%%%%%
%
For the measure of interest we get:
%%%%%%%%%%%%%%%%%%%%%%%%%%
\begin{equation}
\mu = \frac{du\,dv}{v^2} = {\rm J(x,y)}\,|c\tau +d|^4\,\frac{dx\,dy}{y^2}\,,
\label{eq:muxy}
\end{equation}
%%%%%%%%%%%%%%%%%%%%%%%
%
where ${\rm J(x,y)}$ is the Jacobian of the change of variables 
of integration:
%%%%%%%%%%%%%%%%%%%%%%%%%%
\begin{equation}
{\rm J(x,y)} 
= \frac{\partial{u}}{\partial{x}}\,
\frac{\partial{v}}{\partial{y}} - \frac{\partial{u}}{\partial{y}}\,
\frac{\partial{v}}{\partial{x}}\,.
\label{eq:Jxy}
\end{equation}
%%%%%%%%%%%%%%%%%%%%%%%%%%%%%
%
Using the expressions for $u$ and $v$ in Eq. (\ref{eq:uvxy}) we find:
%%%%%%%%%%%%%%%%%%%%%%%%%%%
\begin{equation}
\frac{\partial{u}}{\partial{x}} = \frac{\partial{v}}{\partial{y}} 
= \frac{(cx + d)^2 - c^2\,y^2}{|c\tau +d|^4}\,,~~
 \frac{\partial{u}}{\partial{y}} = 
-\,\frac{\partial{v}}{\partial{x}}
= \frac{2c\,(cx+d)\,y}{|c\tau +d|^4}\,.
\label{eq:dudv}
\end{equation}
%%%%%%%%%%%%%%%%%%%%%%
%
Thus,
%%%%%%%%%%%%%%%%%%%%%%%%%%
\begin{equation}
{\rm J(x,y)} = |c\tau + d|^{-4}\,.
\label{eq:Jxyf}
\end{equation}
%%%%%%%%%%%%%%%%%%%%%%%%%
%
Inserting the expression for ${\rm J(x,y)}$ in Eq. (\ref{eq:muxy}) we obtain:
%%%%%%%%%%%%%%%%%%%%%%%%%%
\begin{equation}
d\mu = \frac{du\,dv}{v^2} = \frac{dx\,dy}{y^2}\,.
\label{eq:muuvxy}
\end{equation}
%%%%%%%%%%%%%%%%%%%%%%%
%

%%%%%%%%%%%%%%%%%%%%%%%%%%
%
\subsubsection{Cusp Modular Forms}
%
%%%%%%%%%%%%%%%%%%%%%%%%%%%%

The integral in Eq. (\ref{eq:IntNy2}) is convergent and the normalisation 
$\overline{{\rm N}}^{(K)}_{\rm Y }$ is well-defined and positive for 
``cusp'' modular forms, i.e., for the modular forms 
such that $|Y^{(K)}(\tau)| \sim O(e^{-2\pi\,y})$ 
when $y \equiv {\rm Im \tau}\rightarrow \infty$ 
\footnote{The $q$-expansions of the holomorphic 
cusp modular forms do not contain  constant $q$-independent 
terms.
}.
Among the considered modular forms these are 
$Y_{\mathbf{\hat{1}'}}^{(3)}(\tau)$, 
$Y_{\mathbf{3}}^{(4)}(\tau)$,
$Y_{\mathbf{\hat{3}},2}^{(5)}(\tau)$ and 
$Y_{\mathbf{\hat{3}'},2}^{(7)}(\tau)$.
We will discuss later a modification of the 
``integral'' normalisation $\overline{N}^{(K)}_{\rm Y }$ 
for the modular forms which do not satisfy the 
``cusp'' condition. These are the majority of modular forms in 
 Eqs. (\ref{eq:k1triplet}) - (\ref{eq:k7}).

 As an example, consider the normalisation of the cusp modular form
$Y_{\mathbf{\hat{1}'}}^{(3)}(\tau) = \sqrt{3} 
(\varepsilon \,\theta^5 - \varepsilon^5\, \theta)$. We will use the 
$q-$expansions of $\varepsilon(\tau)$ and  $\theta(\tau)$ 
given in Eq. (\ref{eq:theta_eps_qexp}). In these expansions 
$|q_4| = {\rm exp}(-\pi y/2) \leq 0.257$ since 
$y \geq \sqrt{3}/2$. Correspondingly, to a very good approximation 
$\theta \cong 1 + 2q^4_4$ and $\varepsilon \cong 2q_4$. 
We can safely neglect the terms 
$\propto |\varepsilon|^{10}$ and 
$\propto {\rm Re(\varepsilon\,2q^4_4\,(\varepsilon^5)^*}$ 
in the expression for $|Y_{\mathbf{\hat{1}'}}^{(3)}(\tau)|^2$
since their contributions to $\overline{{\rm N}}^{(3)}_{\rm \hat{1}' }$ 
is negligible. 
For the normalisation squared of the considered modular form
we get up to corrections $O(10^{-3})$
%%%%%%%%%%%%%%%%%%%%%%%
\begin{equation}
(\overline{N}^{(3)}_{\rm \hat{1}'})^2 \cong 
\int_{-\,\frac{1}{2}}^{\frac{1}{2}}dx \int_{\sqrt{1-x^2}}^{\infty}dy\,\, 
8\,y \left [12\,e^{-\,\pi y} - 144\,e^{-\,3\pi y} \cos(2\pi x)\right]\,.
\label{eq:Nhat1prime3}
\end{equation}
%%%%%%%%%%%%%%%%%%%%%%%%%%%%%
%
The lower limit of the integration over $y$, $\sqrt{1-x^2}$, 
corresponds to the arc bounding the fundamental domain from below,  
on which $x^2 + y^2 = 1$. 
Performing the integration we find:
%%%%%%%%%%%%%%%%%%%%%%%%%%
 \begin{equation}
 \overline{{\rm N}}^{(3)}_{\rm \hat{1}'} \cong  1.38\,.
\label{eq:Nhat1primew3}
 \end{equation}
%%%%%%%%%%%%%%%%%%%%%%
%
It follows form this result that the introduced  
``global'' normalisation of $Y_{\mathbf{\hat{1}'}}^{(3)}(\tau)$
is not precisely 1, but 
is close to 1 and thus cannot alter significantly
the value of $Y_{\mathbf{\hat{1}'}}^{(3)}(\tau)$ at $\tau = \tau_v$.
As can be shown, the same conclusion holds
for the ``global'' normalisations of 
the other cusp modular forms $Y_{\mathbf{3}}^{(4)}(\tau)$,
$Y_{\mathbf{\hat{3}},2}^{(5)}(\tau)$ and $Y_{\mathbf{\hat{3}'},2}^{(7)}(\tau)$
considered by us.

%%%%%%%%%%%%%%%%%%%%%%%%%%
%
\subsubsection{Non-Cusp Modular Forms}
%
%%%%%%%%%%%%%%%%%%%%%%%%%%%%

  A method of normalisation of the holomorphic modular forms 
such that $|Y^{(K)}_{{\bf r}}(\tau)|$ does not 
decrease as $O(e^{-2\pi\,y})$ 
when $y \equiv {\rm Im \tau}\rightarrow \infty$ 
was proposed by D. Zagier \cite{DZ1982}.
It represents the following  modification 
of the ``global'' normalisation defined 
in Eq. (\ref{eq:IntNy2}):
%%%%%%%%%%%%%%%%%%%%%%%%%%%%%%%%
\begin{equation}
\begin{aligned}
(\overline{{\rm N}}^{(K)}_{\rm Y R})^2
& = \lim_{T\to \infty} 
\left ( \int_{D_T} \int (N^{(K)}_{\rm Y} (\tau,\tau^*))^2\,\frac{dx dy}{y^2}
- |a_0|^2 {\rm \frac{2^{K}}{K-1}\, T^{K-1} } \right )
\\
& \equiv 
\lim_{T \to \infty} 
\left (\int_{D_T} \int \sum_i |Y^{(K)}_{{\bf r}i}(\tau)|^2 (2\,y)^{\rm K} \frac{dx\,dy}{y^2} 
- |a_0|^2 {\rm \frac{2^{K}}{K-1}\,T^{K-1} }\right )\,.
\label{eq:IntNy2DZ}
\end{aligned}
\end{equation}
%%%%%%%%%%%%%%%%%%%%%%%%%%%%%%%%%%%%%%
%
Here $T$ is a real constant, 
the region $D_T$ is the lower part of the fundamental domain 
limited from above by the line $y = T$, $T > 1$, 
and $|a_0|^2$ is the constant $q$-independent term in the 
$q$-expansion of $\sum_i |Y^{(K)}_{{\bf r}i}(\tau)|^2$.
The mathematical justification of the normalisation 
(\ref{eq:IntNy2DZ}) and the theorem on which it is based 
can be found in Ref. \cite{DZ1982} and we are not going to 
reproduce them here.
The integration in (\ref{eq:IntNy2DZ}) 
is in the intervals $-\,0.5 \leq x \leq 0.5$ 
and $\sqrt{1-x^2} \leq y \leq T$. The limit 
$T\rightarrow i\infty$ is taken after the 
integration over the variables $x$ and $y$ 
is performed 
\footnote{The procedure of normalisation of the 
``non-cusp'' modular forms proposed in 
\cite{DZ1982} formally resembles the procedure 
of renormalisation of the amplitudes of processes 
in quantum field theory by which the infinite terms are removed.
Actually, considering the inner product of two modular forms 
$f(z)$ and $g(z)$ of weight $k$ neither of which is a cusp form,
the author of \cite{DZ1982} writes that the modular invariant 
function $F(z) = y^kf(z) \overline{g}(z)$ is 
``renormalisable'' and that one can define the inner product 
of $f(z)$ and $g(z)$ as its ``renormalisable integral''
over the fundamental domain $D$. As should be clear, 
we are considering the case of $g(z) \equiv f(z)$. 
}.

In the case of the cusp modular forms $a_0 = 0$,
and since $D_T \rightarrow D$ when $T \rightarrow i\infty$,
the normalisation defined in Eq. (\ref{eq:IntNy2DZ}) 
reduces to that defined in Eq.  (\ref{eq:IntNy2}).

We note also that the method of normalisation 
of ``non-cusp'' modular forms was derived in \cite{DZ1982} 
under the assumption that the modular form weights 
satisfy $K \geq 2$. We assume that in the case of modular forms 
with weight $K = 1$ the term $|a_0|^2 {\rm 2^K T^{K-1}/(K-1) }$ 
in Eq. (\ref{eq:IntNy2DZ}) 
should be replaced by $|a_0|^2\,{\rm 2^K\, ln T}$.

Finally, it follows from Eq. (\ref{eq:IntNy2DZ}) that 
$(\overline{{\rm N}}^{(K)}_{\rm Y R})^2$ is a real number 
but is not necessarily positive-definite \cite{DZ1982}.
In the cases when $(\overline{{\rm N}}^{(K)}_{\rm Y R})^2 < 0$ 
we propose to use the absolute value  
$|\overline{{\rm N}}^{(K)}_{\rm Y R}|$ as a ``global'' 
normalisation of $Y^{(K)}_{\bf r}(\tau)$. 

 As an example of the application of the method 
defined in Eq. (\ref{eq:IntNy2DZ})
we show next the normalisation of the level 4 weight 3 
``non-cusp'' modular form $Y^{(3)}_{\hat{3}'}(\tau)$ 
whose expression in terms of the functions 
$\theta(\tau)$ and $\varepsilon(\tau)$
is given in Eq. (\ref{eq:k3}). 
We recall that to a very good approximation 
$\theta \cong 1 + 2q^4_4$ and $\varepsilon \cong 2q_4$ 
as $|q_4| = {\rm exp}(-\pi y/2) \leq 0.257$ since 
$y \geq \sqrt{3}/2$. Up to corrections of $O(10^{-2})$, we find:
%%%%%%%%%%%%%%%%%%%%%%%%%%%%%%%%%%%
\begin{equation}
\sum_i |Y^{(3)}_{\hat{3}' i}(\tau)|^2 \cong 
\frac{1}{4}
\left [1 + 2048\,e^{-3\pi y} 
+ \left (144 + 120\,\cos(2x\pi)\right)\,e^{-2\pi y}\right ]\,. 
\label{eq:N2hat3p}
\end{equation}
%%%%%%%%%%%%%%%%%%%%%%%%%%%%%%%%%%%%%
%
Inserting the derived  expression 
for $\sum_i |Y^{(3)}_{\hat{3}' i}(\tau)|^2$ in Eq. (\ref{eq:IntNy2DZ}) 
and performing the integration we find for 
the ``global'' normalisation factor for 
 $Y^{(3)}_{\hat{3}'}(\tau)$ a value close to 1:
%%%%%%%%%%%%%%%%%%%%%%%%%%%%%%%%
\begin{equation}
|\overline{{\rm N}}^{(3)}_{\rm \hat{3}'  R}| \cong 0.9\,. %0.86\,.
\label{eq:NY3hat3pR}
\end{equation}
%%%%%%%%%%%%%%%%%%%%%%%%%%%%%%%%%%
%
% We estimate an uncertainty of $\sim 6\%$ in the quoted value.
As in the case of the example of  normalisation of the cusp 
form $Y_{\mathbf{\hat{1}'}}^{(3)}(\tau)$ we have considered, 
the global normalisation factor of the ``non-cusp'' 
form $Y^{(3)}_{\hat{3}'}(\tau)$ is close to 1 and therefore, 
when applied, it cannot change significantly 
the magnitude of the components of $Y_{\mathbf{\hat{1}'}}^{(3)}(\tau)$ 
at $\tau = \tau_v$.

 The ``global'' normalisations of other modular forms, 
which are holomorphic functions of the modulus $\tau$  
and do not satisfy the cusp condition, can be calculated 
in a similar way.

%%%%%%%%%%%%%%%%%%%%%%%%%%%%%%%%%%%%%%%%%%%%%%
%
\section{Summary}
%
%%%%%%%%%%%%%%%%%%%%%%%%%%%%%%%%%%%%%%%%%%%%%%

We have discussed the normalisation 
of the modular forms 
which  play a crucial role in the modular invariance approach 
to the flavour problem. 
In this approach the elements of the fermion Yukawa 
couplings and mass matrices in the Lagrangian of the theory 
are modular forms of a certain level \(N\) and 
weights $K$. As like the fermion (matter) fields, the 
modular forms of   level \(N\)
have specific transformation properties 
under the action of the inhomogeneous (homogeneous) 
modular symmetry group $\Gamma \equiv PSL(2,\mathbb{Z})$
($\Gamma^\prime \equiv SL(2,\mathbb{Z})$).  
They also furnish irreducible representations of the
the inhomogeneous (homogeneous) finite modular 
group $\Gamma^{(\prime)}_{\rm N}$, which describes  
the flavour symmetry. The modular forms are holomorphic 
functions of a single complex scalar field -- the modulus $\tau$. 
When  $\tau$ develops a VEV, $\tau_v$, the modular forms 
in the fermion mass matrices get fixed and certain 
flavour structure arises.
 
It is a well known fact that the modular forms 
furnishing irreducible representations of the 
finite modular groups are determined up to a constant.
Usually these normalisation constants 
are absorbed in the constant parameters 
which multiply each modular form 
present in the fermion mass matrices. 
Since the normalisation of the modular forms 
is arbitrary, this makes the specific values of the 
constant parameters, obtained in a given model 
by statistical analysis of the
description of the relevant experimental data 
by the model, of not much physical meaning. 
Moreover, it also makes the comparison of 
models which use different normalisations  
of the modular forms ambiguous.
The problem of normalisation of the 
modular forms can be particularly acute in 
modular flavour models in which the 
charged lepton and quark mass hierarchies 
are obtained without fine-tuning  
the constant parameters present in the 
respective fermion mass matrices, 
where the constants have to be of the same 
order in magnitude.

  We have discussed two possible modular invariant normalisations 
of the modular forms $Y^{(K)}$ of interest 
-- a ``local'' one at $\tau = \tau_v$ (Eq. (\ref{eq:Ny})), 
$N^{(K)}_Y (\tau_v)$,  
and a ``global'' one, $\overline{{\rm N}}^{(K)}_{\rm Y}$, 
involving an integration of 
$(N^{(K)}_Y (\tau))^2$ over the 
fundamental domain $D$ of the modular group 
over the modular invariant hyperbolic 
measure (or volume form) $d\mu(\tau) = dx\,dy/y^2$
(Eq. (\ref{eq:IntNy2})). 
Both normalisations do not depend on the basis employed 
for the generators of the finite modular group 
$\Gamma^{(\prime)}_{\rm N}$ describing the flavour symmetry.
We have considered the simple case when the 
Higgs fields present in the Yukawa couplings are singlets with 
zero weights of the considered finite modular 
group describing the flavour symmetry. However, 
our results can be easily generalised to more 
sophisticated cases including also flavon fields.
In the considered  case of singlet Higgs fields, 
the local normalisation of a given modular 
form $Y^{(K)}_{\bf r}$, present in a modular invariant 
fermion mass term (\ref{eq:bilinear}), {\bf r} being the representation 
of the respective finite modular (flavour) group,
effectively coincides with the Euclidean norm 
of the form at $\tau = \tau_v$,
$(\sum_i |Y^{(K)}_{{\bf r}i}(\tau_v)|^2)^{\frac{1}{2}}$, 
after one performs the necessary 
renormalisation of the fermion fields  
in order to get canonical fermion kinetic terms
in the Kahler potential of the theory.
In this case the fermion normalisation factors 
do not appear in the fermion mass matrices. 

For the cusp modular forms, the global normalisation 
$\overline{{\rm N}}^{(K)}_{\rm Y}$ (as we have learned in 2023) 
is based on the Petersson inner product of two 
modular forms \cite{Pet1932},
while in the case of ``non-cusp'' (holomorphic) modular  
forms the global normalisation $\overline{{\rm N}}^{(K)}_{\rm YR}$
(Eq. (\ref{eq:IntNy2DZ})) is based on the Zagier inner 
product of two modular forms \cite{DZ1982}, 
which represents a ``renormalised'' 
modification of the Petersson inner product. 
The square of  $\overline{{\rm N}}^{(K)}_{\rm YR}$ 
calculated following the Zagier recipe 
is real, but is not 
% guaranteed to be 
necessarily positive-definite. 
In the cases of $(\overline{{\rm N}}^{(K)}_{\rm YR})^2 < 0$,
we have proposed to use $|\overline{{\rm N}}^{(K)}_{\rm YR}|$ 
as a global normalisation of the non-cusp holomorphic 
modular forms.

  We find that the local normalisation of the cusp modular forms 
at $\tau = \tau_v$, present in the mass term  (\ref{eq:bilinear}),  
which coincides with the Euclidean norms 
of the forms at  $\tau = \tau_v$, 
can change drastically their magnitude if the Euclidean norm 
is much smaller than 1.
As we have shown, this can have important implications 
for the description of observables such as the fermion masses,
if cusp forms are present in the fermion (quark, charged lepton)
mass matrices in the modular flavour models. 
In the case of the non-cusp holomorphic modular forms,
their local normalisations are $\sim 1$ and essentially 
have no effect on the magnitude of the forms at $\tau = \tau_v$.

 In what concerns the global normalisation of 
cusp and non-cusp holomorphic modular forms, we have found that 
their respective normalisation factors are close to 1 
and can change somewhat, but not in a significant way, 
the magnitude of the forms.
In contrast to the case of the local normalisation, 
the fermion normalisation factors have to be 
taken into account in the fermion mass matrices 
if one uses the global normalisation of the modular forms.

Our results indicate that the modular invariant global 
normalisation of the holomorphic modular forms 
present in the fermion Yuakawa couplings and 
mass matrices in modular flavour models seems better 
suited for the formalism of the modular invariance 
approach to the flavour problem.
  
\vspace{-0.6cm}
%%%%%%%%%%%%%%%%%%%%%%%%%%%%%%%%%%%%%%%%%%%%%%
\section*{Acknowledgements}
%%%%%%%%%%%%%%%%%%%%%%%%%%%%%%%%%%%%%%%%%%%%%%

The author would like to thank P. Novichkov, J. Penedo,
A. Titov and  A. M. Kidambi for very useful discussions, 
and J. Penedo for checking the results 
for the global normalisations of the modular forms 
$Y_{\mathbf{\hat{1}'}}^{(3)}(\tau)$ and $Y^{(3)}_{\hat{3}'}(\tau)$ 
and correcting my result for that of $Y_{\mathbf{\hat{1}'}}^{(3)}(\tau)$.
This work was supported in part by the European Union's Horizon 2020 research 
and innovation programme under the Marie Sklodowska-Curie
% Sk\l{}odowska-Curie 
grant agreement 
No.~860881-HIDDeN, by the Italian INFN program on Theoretical Astroparticle 
Physics and by the World Premier International Research Center Initiative 
(WPI Initiative, MEXT), Japan.

\vskip 0.5cm
%%%%%%%%%%%%%%%%%%%%%%%%%%%%%%%%%%%%%%

\end{document}